\documentclass{JAC2003}
\usepackage{graphicx}
\usepackage{booktabs}

\begin{document}
\title{LARGE ENERGY ACCEPTANCE DOGLEG FOR THE XFEL INJECTOR
\vspace{-0.5cm}}

\author{V. Balandin, W. Decking, N. Golubeva\thanks{nina.golubeva@desy.de} \\
DESY, Hamburg, Germany}

\maketitle

\begin{abstract}
The option to install two injectors is foreseen at the European XFEL Facility~\cite{XFEL}. 
The injectors will be located on top of each other in the same building,
both with the offset of 2.75 m with respect to the main linac axis.
The translation system (dogleg) from the injector 
axis to the main linac axis has
to fulfill
very tight requirements of the chromatic
properties, because the energy chirp required for
the downstream bunch length compression in magnetic chicanes will
be created upstream in the injector linac.
In this paper we present such an large energy acceptance dogleg 
and discuss the optical principles which form the basis of its design.
\end{abstract}

\section{DESIGN RECIPES FOR LARGE ENERGY ACCEPTANCE DOGLEG}

We will consider a parallel beam translation system (dogleg) where a bend 
magnet block (arc) which is symmetric about the horizontal midplane $y = 0$ 
is followed by its rotated by $180^\circ$ about the longitudinal axis image.

In order to design a dogleg which does not give rise to the unacceptable emittance 
dilution due to chromatic effects one has to control both, dogleg nonlinear
dispersions and dogleg chromatic focusing properties, and that can be done
using the following observations:

Let us assume that the arc transport matrix is free from the linear dispersions
and its horizontal focusing part is equal to the two by two identity matrix.
Then in the dogleg transfer map the second order dispersions are automatically
canceled. If we will add to these assumptions the requirement that the arc map 
is a second order achromat, we will obtain dogleg which is a second order achromat 
by itself and in which the first nonzero dispersions are at least of fifth order.

These observations were used during design of a beam transport system for the TESLA 
X-ray Facility and allowed to work out dogleg with the energy acceptance which 
definitely is not smaller than $\pm 10 \%$ ~\cite{DistrTESLA, TESLA}. The sextupoles 
used in that dogleg are essential for achieving such large energy acceptance and play 
a twofold role. They are responsible for both, control of chromatic focusing aberrations 
by making the arc to be a second order achromat and for absence of nonlinear dispersions 
up to fifth order. 

Nevertheless, as concerning suppression of nonlinear dispersions 
alone, then the tuning of the arc to be a complete second order achromat is not necessary. 
It is sufficient to make it a second order achromat only with respect to the bending plane
(horizontal) motion (i.e. to make the horizontal components of the arc map free from the 
second order chromatic and geometric aberrations on the manifold $y = p_y = 0$) and thus 
to reduce the number of sextupoles required for cancellation of third and fourth order 
dispersions by a factor of two. The additional possibilities for dispersion cancellation
we will obtain, if we will assume that the arc is constructed by a repetition of $n$ 
identical cells ($n > 1$) with the arc horizontal focusing matrix equal to the two by two 
identity matrix and with the cell horizontal focusing matrix not equal to the two by two 
identity matrix (which guarantees that the arc transport matrix is automatically free from 
the linear dispersions). In this case let us summarize the rules for the dispersion
suppression as follows:

\begin{itemize}
\item
Without any additional assumptions the second order dogleg dispersions are equal to zero.
\item
If the arc cell is free form the linear dispersions, then the second and the third order 
dogleg dispersions are equal to zero.
\item
If the arc map is a second order achromat with respect to the horizontal motion, then
the second, the third and the fourth order dogleg dispersions are equal to zero.
\item
If the arc map is a second order achromat with respect to the horizontal motion and the arc 
cell is free form the linear dispersions, then the second, the third, the fourth and the fifth
order dogleg dispersions are equal to zero.
\end{itemize}

One sees from the above list that it is not a big problem to construct dogleg without sextupoles 
which, nevertheless, is free from first, second and third order dispersions, but what is it 
possible to do in this case (or in the case when the arc map is a second order achromat only 
with respect to the horizontal motion) in order to control chromatic focusing aberrations? 
One possible way is to employ concept of apochromatic focusing which, though have been developing 
mostly for straight drift-quadrupole systems~\cite{ApIPAC10, ApIPAC11}, can be generalized  
on the systems with bending and sextupole magnets included.\footnote{The theory of apochromatic 
focusing states that for every drift-quadrupole system there exists an unique set of
Twiss parameters, which will be transported through that system without first order chromatic 
distortions. The most interesting for us in this paper statement of this theory is
the statement concerning apochromatic Twiss parameters of periodic systems ~\cite{ApIPAC11},
which is based on averaging and can be extended to include bend magnet systems with sextupoles.}

\section{SOLUTIONS FOR THE XFEL INJECTOR DOGLEG}

In this section we will present three solutions for
the XFEL injector dogleg designed according to the discussed
above optical principles.

\begin{figure}[h]
    \centering
    \includegraphics*[width=80mm]{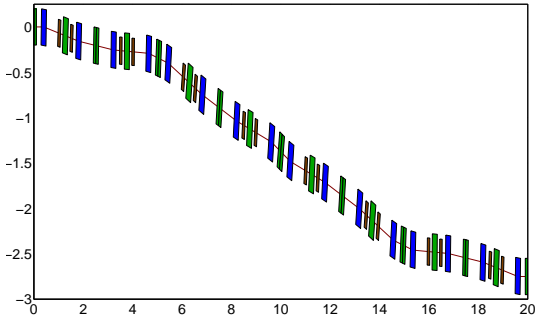}
    \vspace{-0.4cm}
    \caption{Overall layout of the dogleg variant one.
    Blue, green and brown colors mark dipole, quadrupole
    and sextupole magnets, respectively.}
    \vspace{-0.1cm}
    \label{fig1}
\end{figure}

\begin{figure}[h]
    \centering
    \includegraphics*[width=75mm]{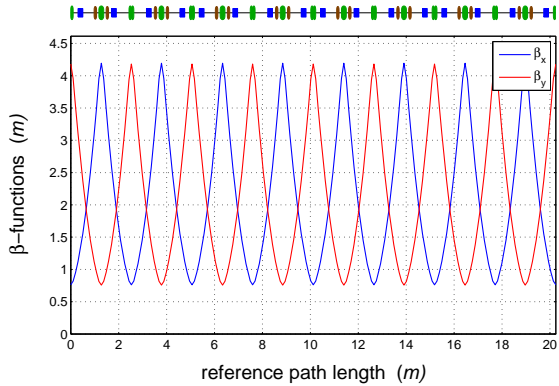}
    \vspace{-0.4cm}
    \caption{Betatron functions along dogleg variant one.}
    \vspace{-0.1cm}
    \label{fig2}
\end{figure}

\begin{figure}[!h]
    \centering
    \includegraphics*[width=75mm]{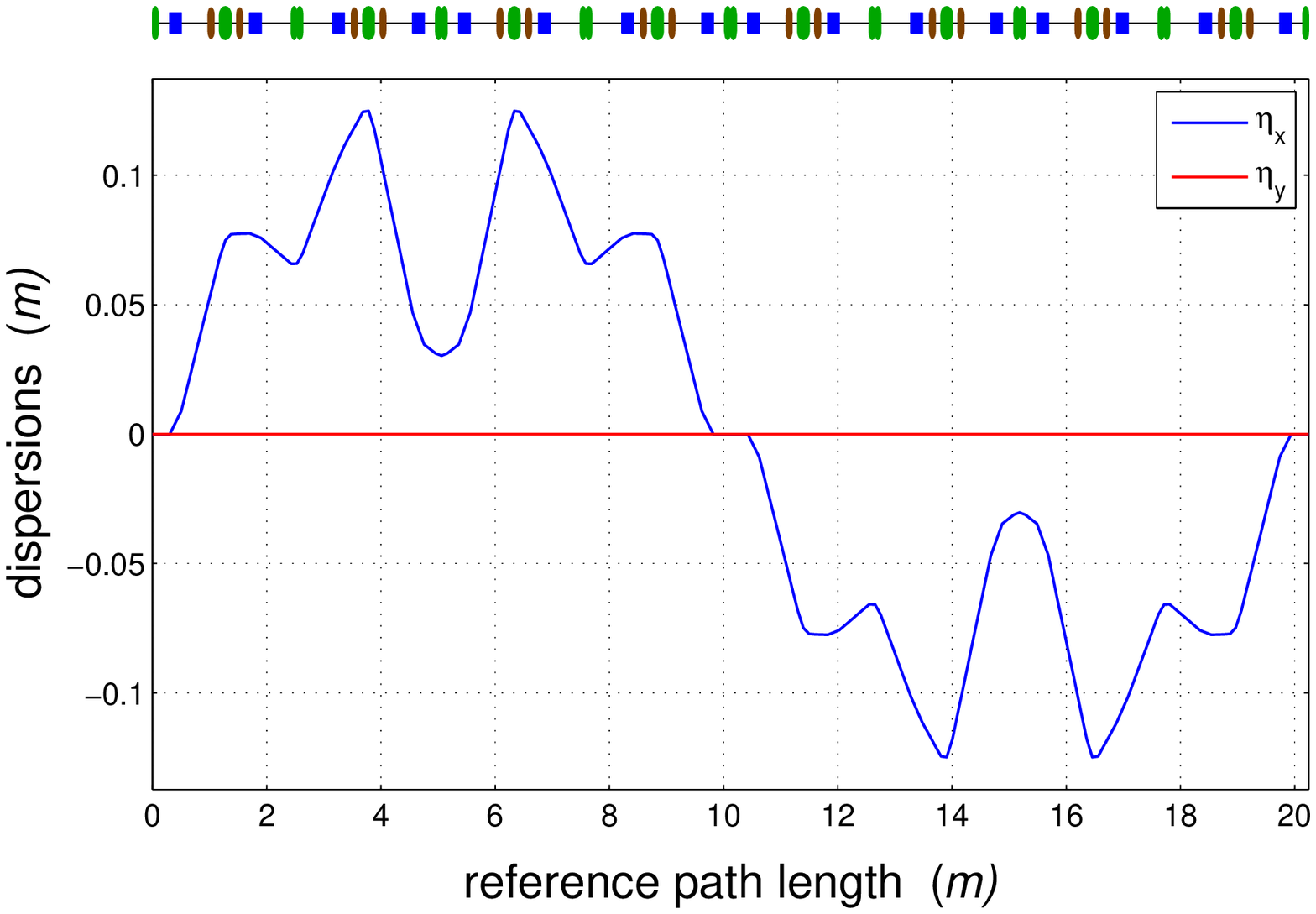}
    \includegraphics*[width=75mm]{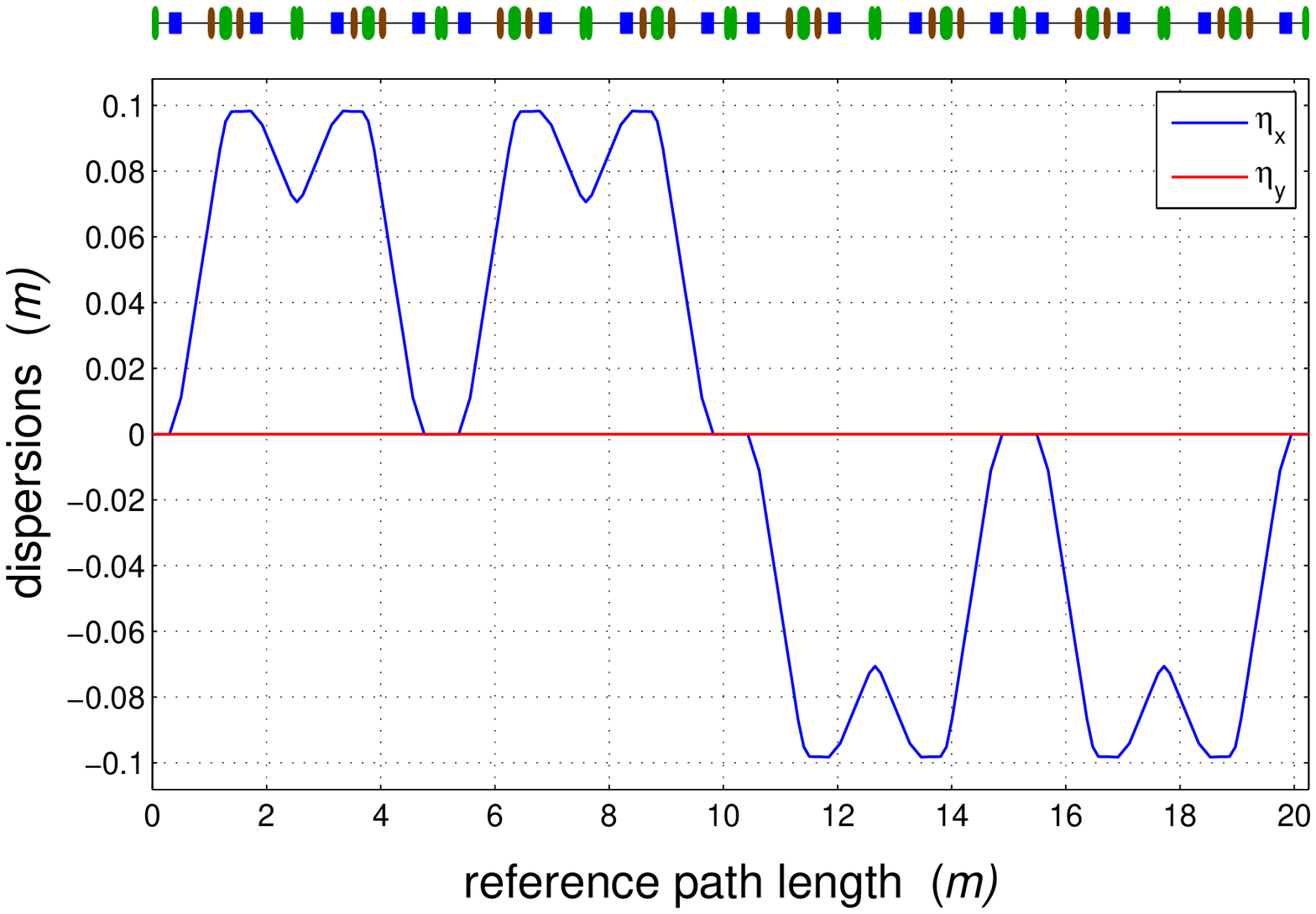}
    \vspace{-0.4cm}
    \caption{Dispersions along doglegs variant one (top) and variant two (bottom).}
    \vspace{-0.4cm}
    \label{fig3}
\end{figure}

The first solution (dogleg variant one) is the dogleg which
originally was designed for the XFEL injector ~\cite{InjLINAC08}.
The second solution (dogleg variant two) is the slight modification of the first
and has better chromatic properties with sextupoles switched off.
The third solution (dogleg variant three) utilizes 
much smaller number of magnets than the first two, does not uses
sextupoles at all, but still has an energy acceptance of the order of
$\pm 3\%$, $\pm 3.5\%$. It is perspective solution and, 
if the practical operations with the first XFEL
injector will show that provided by it energy acceptance is sufficient,
then probably this dogleg will be realized for the usage with the second
XFEL injector.

The doglegs variant one and variant two use the same number of magnets and
have very similar layouts and betatron functions (Fig.1 and Fig.2). 
Arcs of both doglegs are first order achromats, i.e. their transport matrices 
are equal to the identity matrix except for the $r_{56}$ element for the dogleg variant two. 
They can be tuned to become second order achromats with respect to the bending plane 
(in this paper, horizontal) motion using two sextupole families
and are constructed as two cell systems, where each cell is mirror
symmetric with respect to its center and has the same arrangement of dipole and
quadrupole magnets as the cell of the arc of the XFEL 
post-linac collimation section ~\cite{ColXFEL}.

What makes these two doglegs different, it is behavior
of their linear dispersions. For the dogleg variant two it is closed already after
one arc cell (Fig.3).
As the result of that even with sextupoles switched off the dogleg variant two  
has an energy acceptance of the order of $\pm 3\%$, $\pm 3.5\%$, while
the dogleg variant one seems to be nonoperational (Fig.4).
The price paid for this dispersion adjustment is that while
the dogleg variant one is first order isochronous beamline with $r_{56} = 0$, 
the dogleg variant two has $r_{56} \approx 3 cm$ (with the same sign as
for the usual four-bend magnetic chicane) and will make slight beam compression
during its transport, but currently it is considered even as an advantage in comparison with
the dogleg variant one. 

Note that with the sextupoles switched on both doglegs show excellent beam transfer
properties (Fig.5).

\begin{figure}[h]
    \centering
    \includegraphics*[width=60mm]{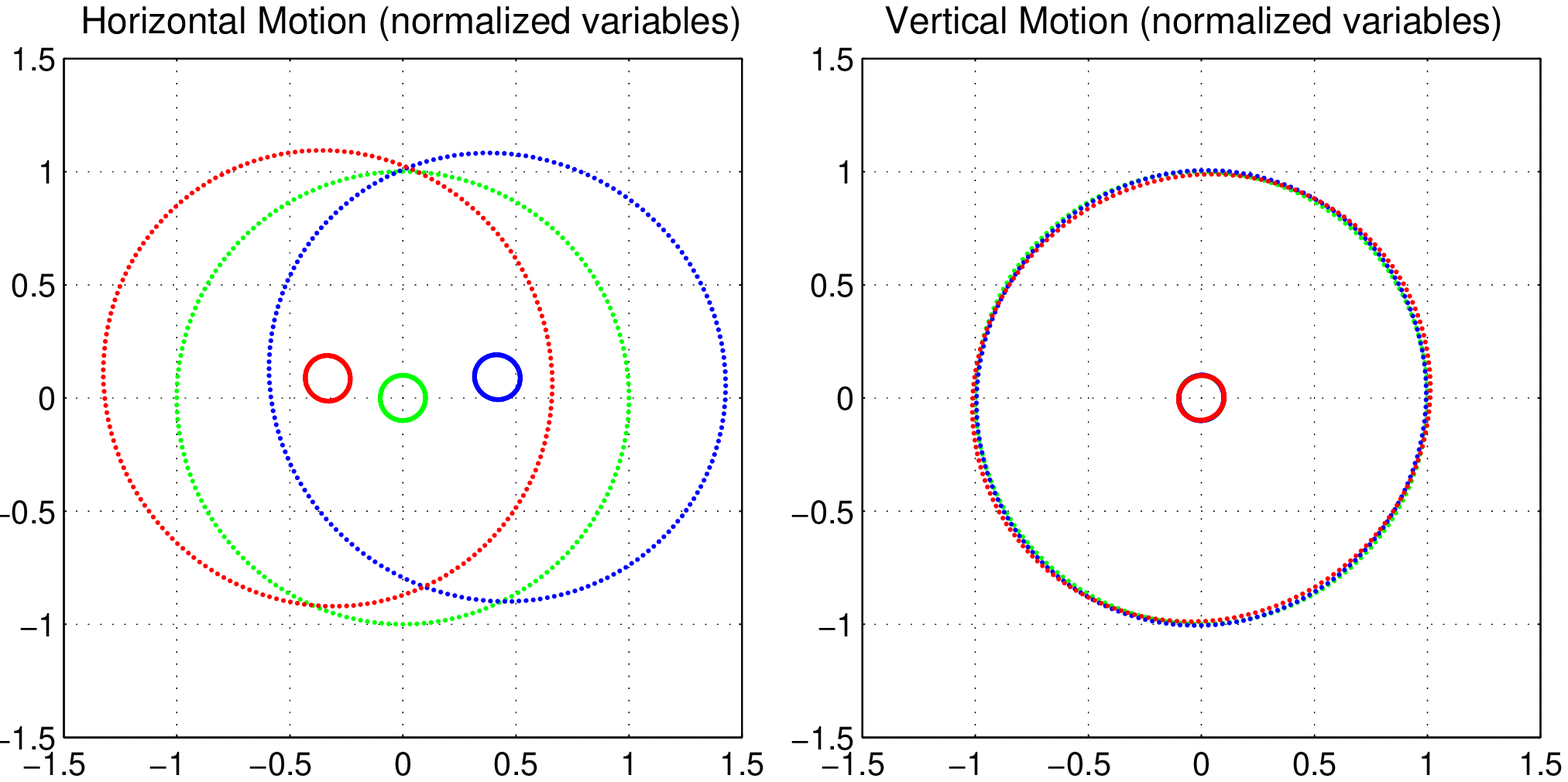}
    \includegraphics*[width=60mm]{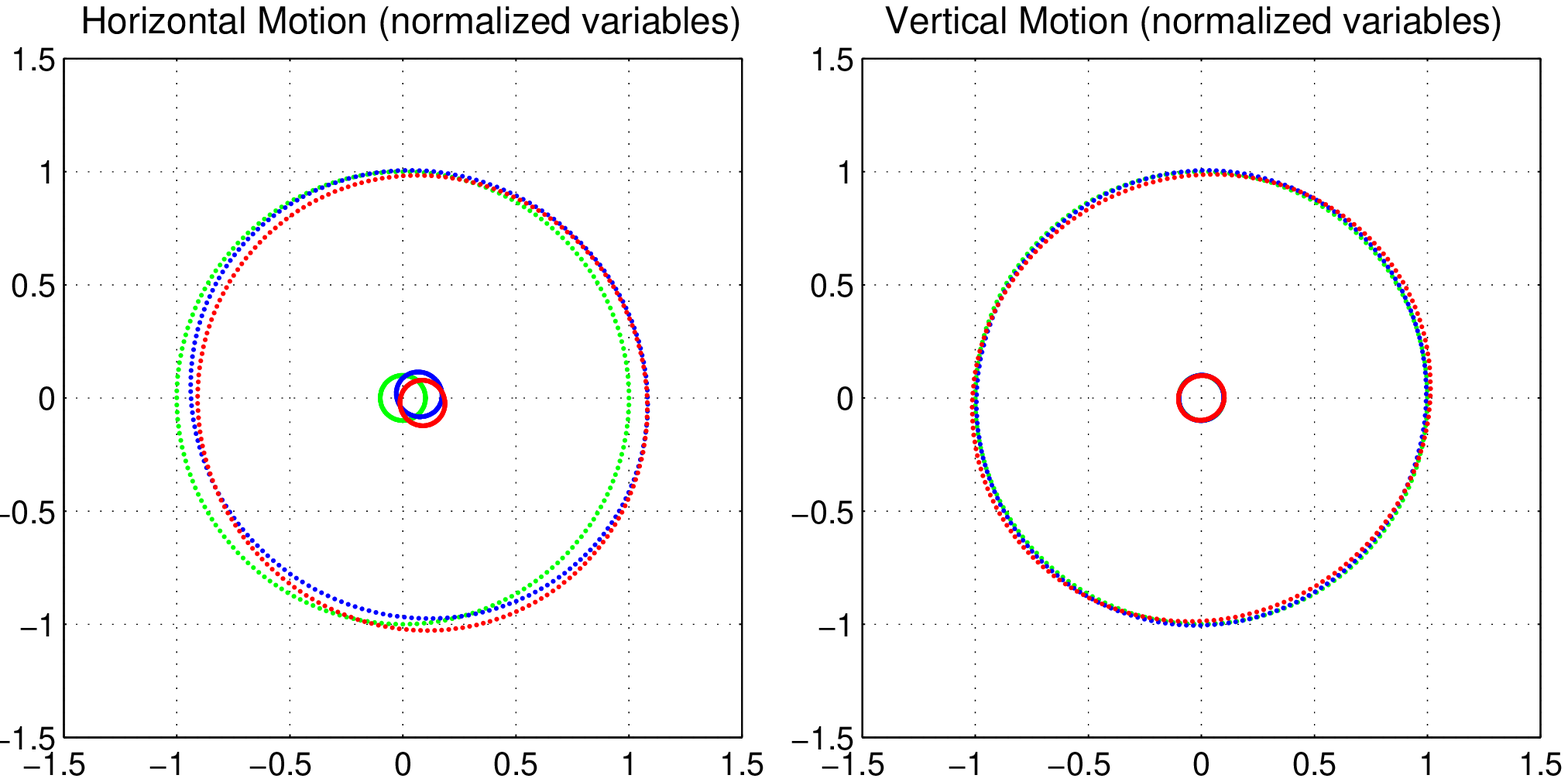}
    \vspace{-0.4cm}
    \caption{Phase space portraits of monochromatic 
    $0.1\sigma_{x,y}$ and $1\sigma_{x,y}$ ellipses
    (matched at the entrance) after tracking through
    the dogleg variant one (top) and the dogleg variant two (bottom). 
    The relative energy deviations are equal to
    $\pm 3\%$.
    Sextupoles are switched off.}
    \vspace{-0.1cm}
    \label{fig4}
\end{figure}

\begin{figure}[!h]
    \centering
    \includegraphics*[width=60mm]{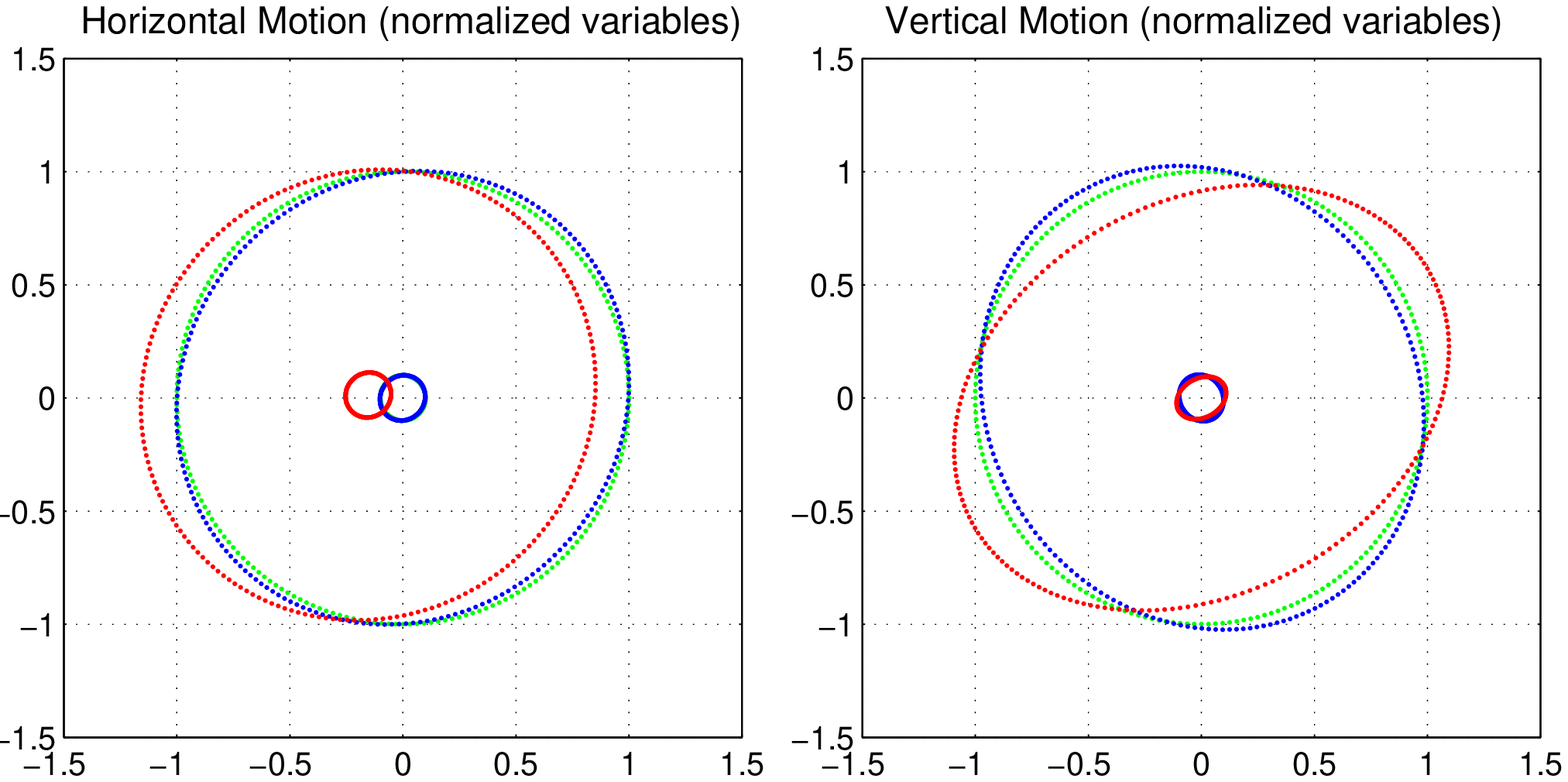}
    \includegraphics*[width=60mm]{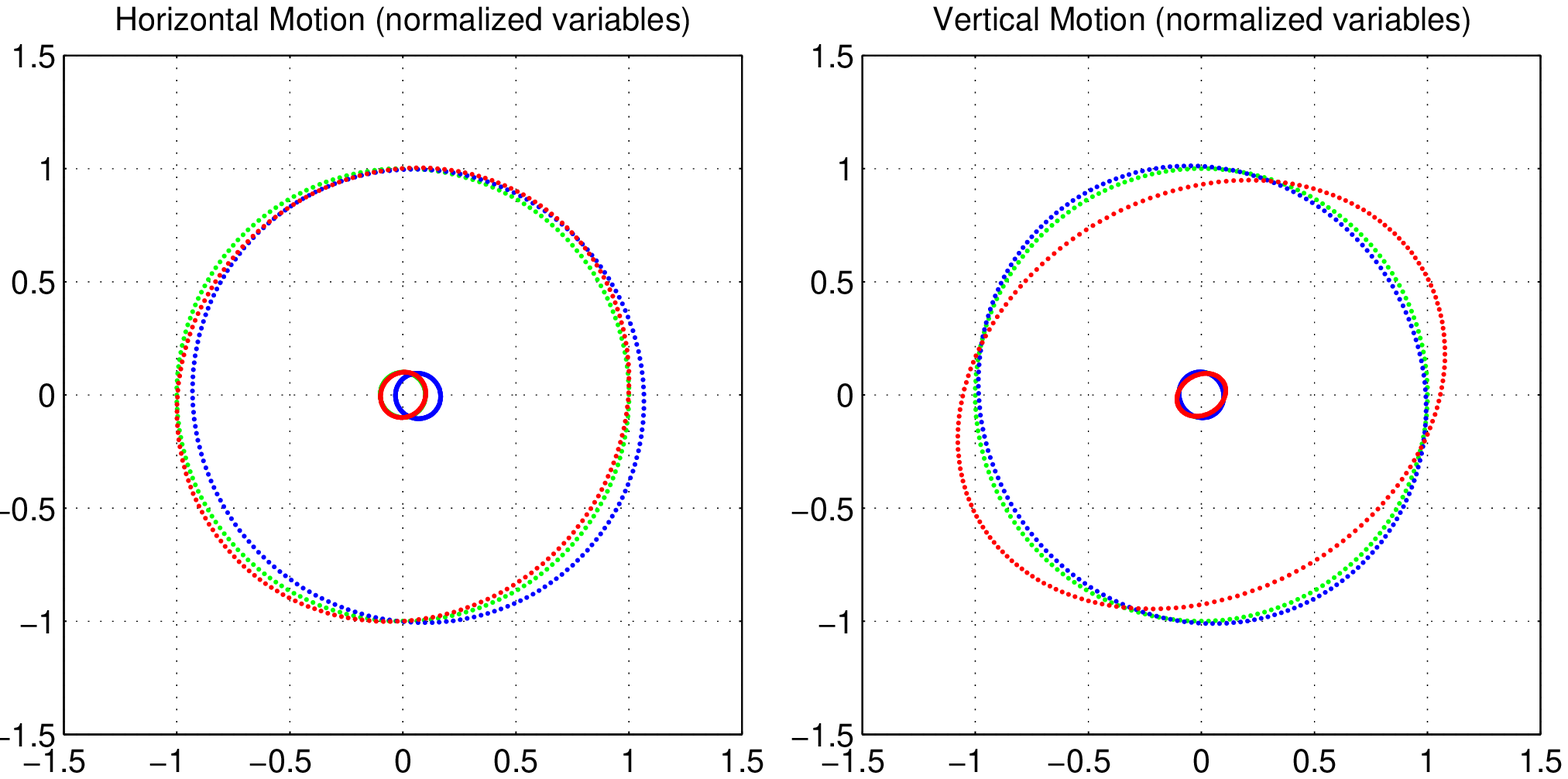}
    \vspace{-0.45cm}
    \caption{Phase space portraits of monochromatic 
    $0.1\sigma_{x,y}$ and $1\sigma_{x,y}$ ellipses
    (matched at the entrance) after tracking through
    the dogleg variant one (top) and the dogleg variant two (bottom). 
    The relative energy deviations are equal to
    $\pm 15\%$.
    Sextupoles are switched on.}
    \vspace{-0.9cm}
    \label{fig5}
\end{figure}

\begin{figure}[!h]
    \centering
    \includegraphics*[width=80mm]{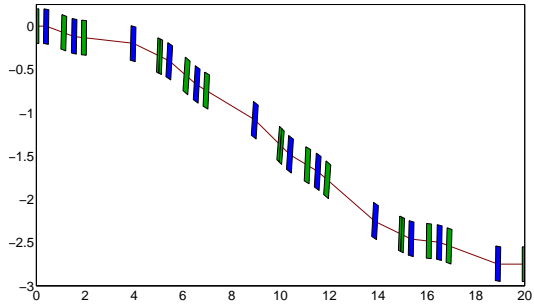}
    \vspace{-0.45cm}
    \caption{Overall layout of the dogleg variant three.
    Blue and green colors mark dipole and quadrupole
    magnets, respectively.}
    \vspace{-0.6cm}
    \label{fig6}
\end{figure}

\begin{figure}[!h]
    \centering
    \includegraphics*[width=75mm]{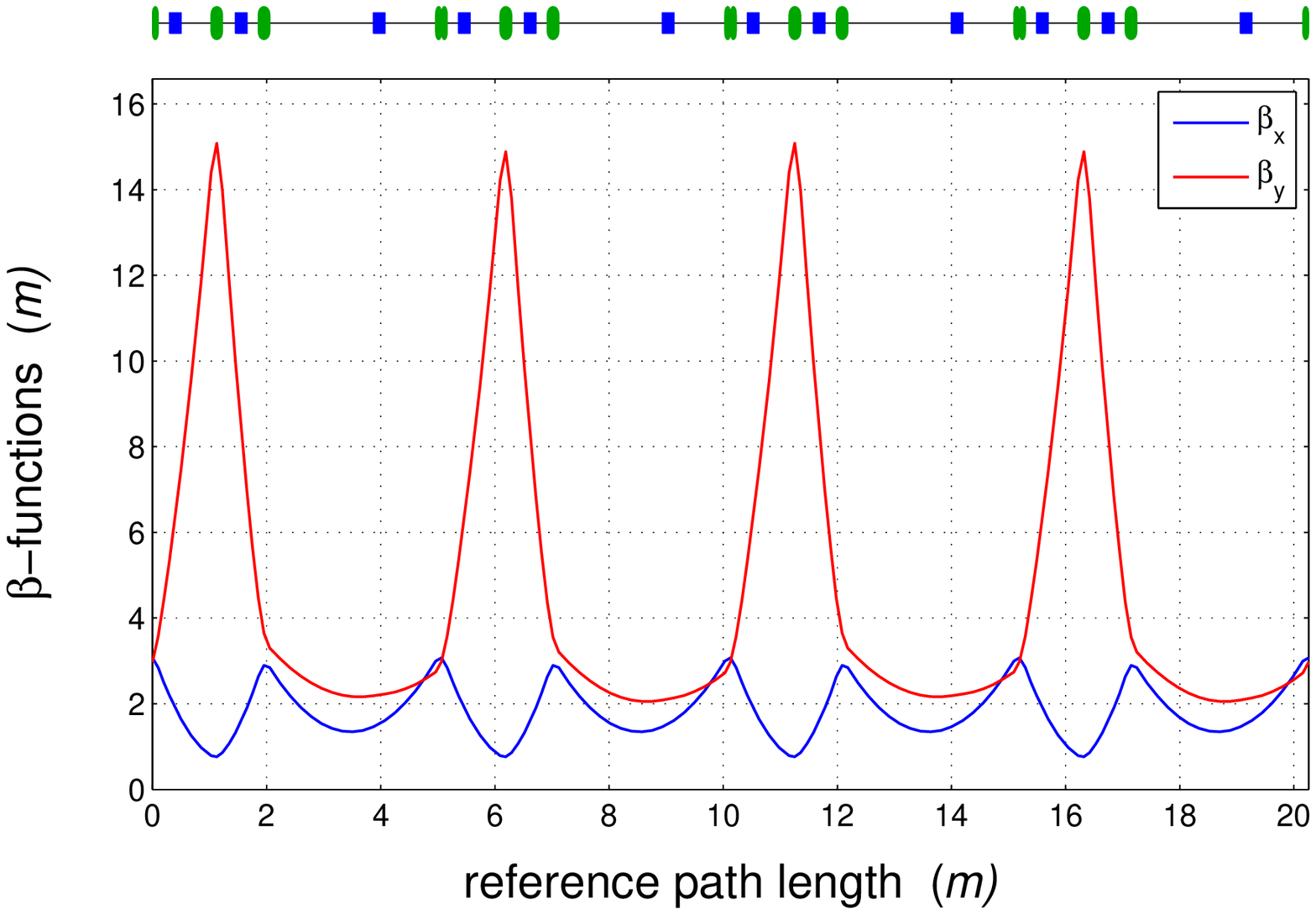}
    \includegraphics*[width=75mm]{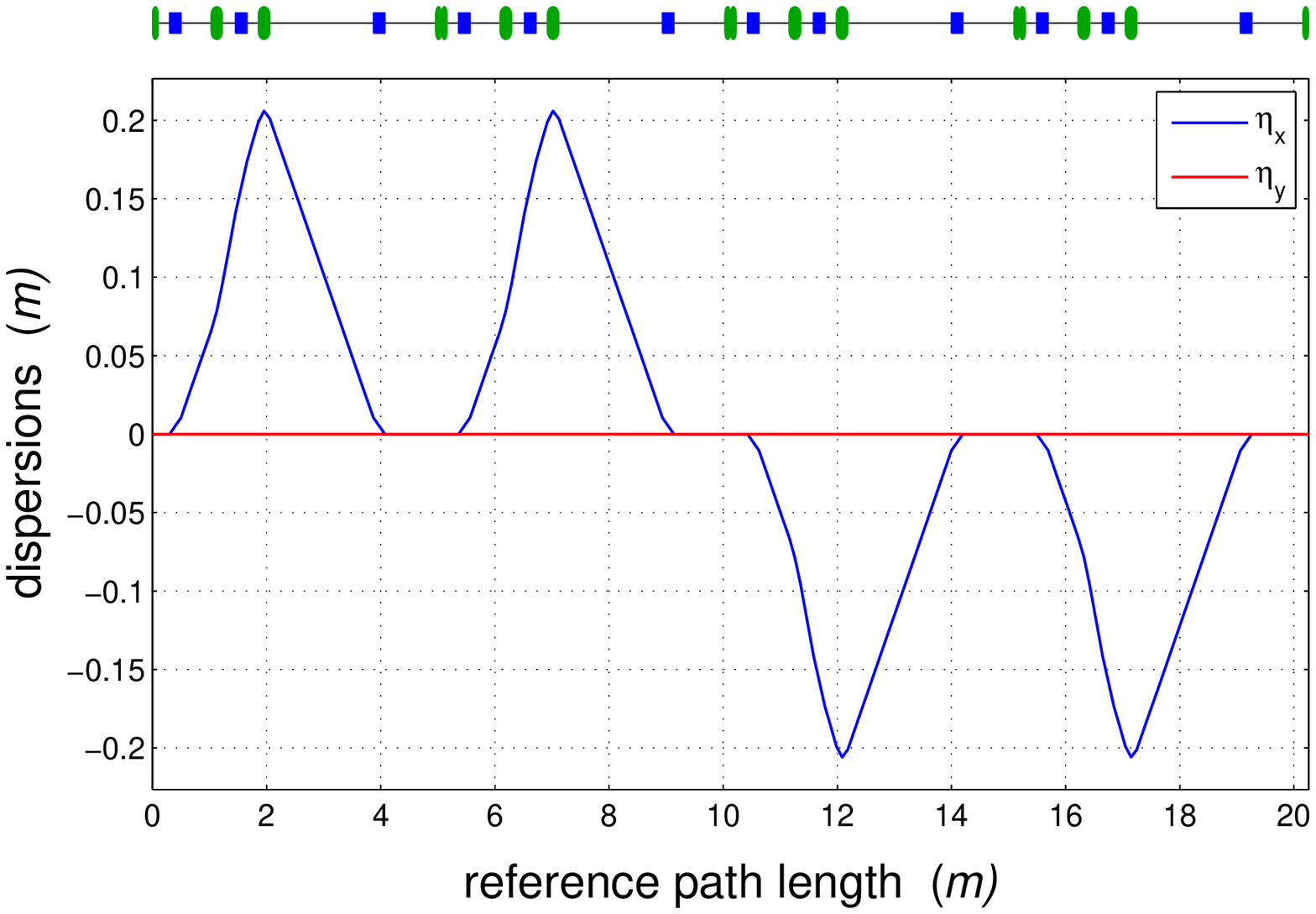}
    \vspace{-0.45cm}
    \caption{Betatron functions and dispersions along dogleg variant three.}
    \vspace{-0.2cm}
    \label{fig7}
\end{figure}

\vspace{0.2cm}

The layout and the Twiss parameters of the dogleg variant three
can be seen at Fig.6 and Fig.7.
This dogleg uses essentially
smaller number of magnets than the first two doglegs presented in this 
paper, which is achieved by reduction of the phase advance
of the vertical motion by a factor of two in comparison with the horizontal motion.
The horizontal focusing part of the arc transfer matrix for this
dogleg is equal to the two by two identity matrix, while its
vertical focusing part is equal to the two by two minus identity matrix.
Similar to the dogleg variant two the dogleg variant three 
will make slight beam compression
during its transport ($r_{56} \approx 4.4 cm$)
and has the linear dispersion which is closed already 
after one arc cell.
The chromatic beam transfer properties of this dogleg
can be seen at Fig.8.

\begin{figure}[!h]
    \centering
    \includegraphics*[width=60mm]{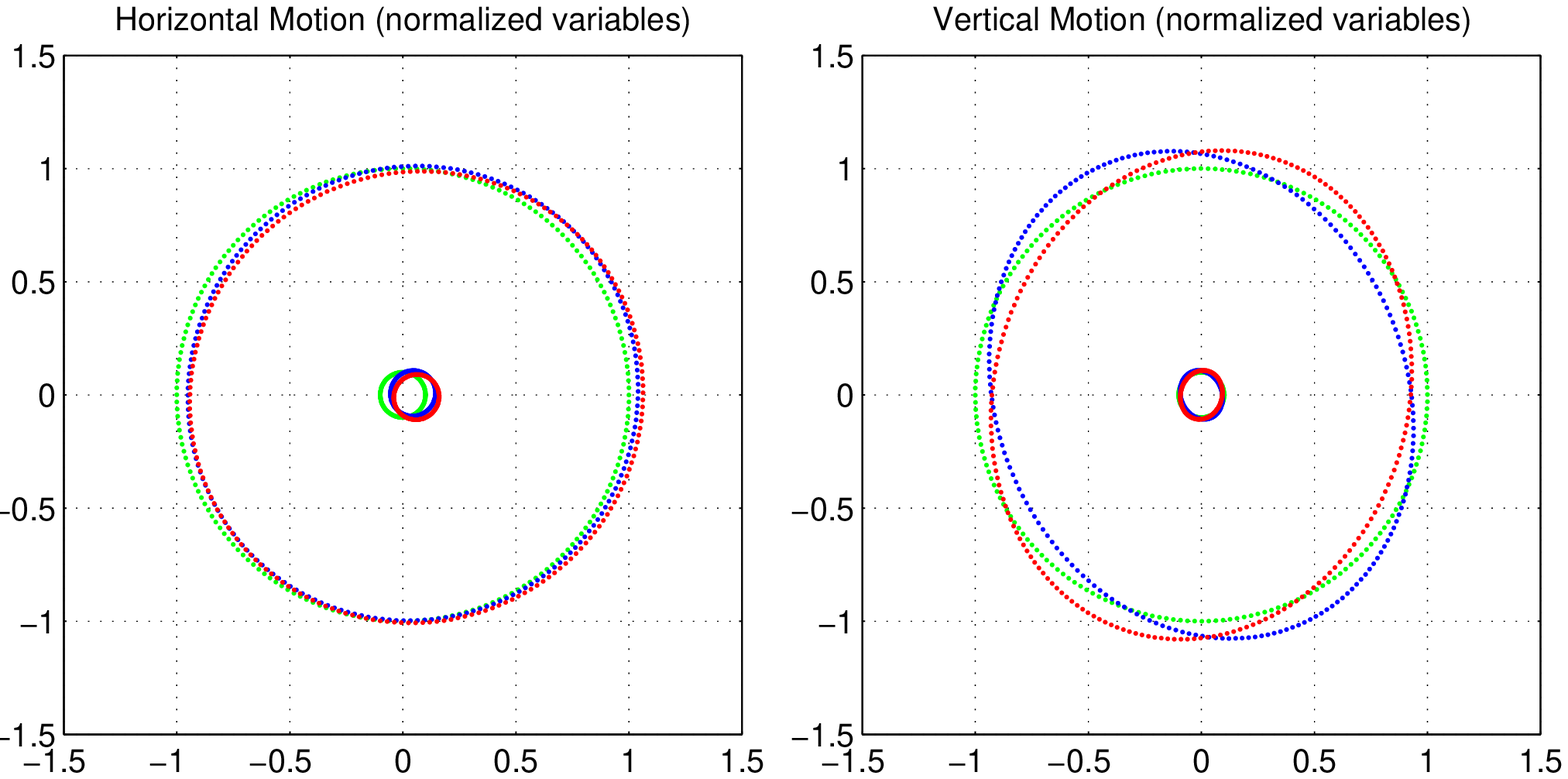}
    \includegraphics*[width=60mm]{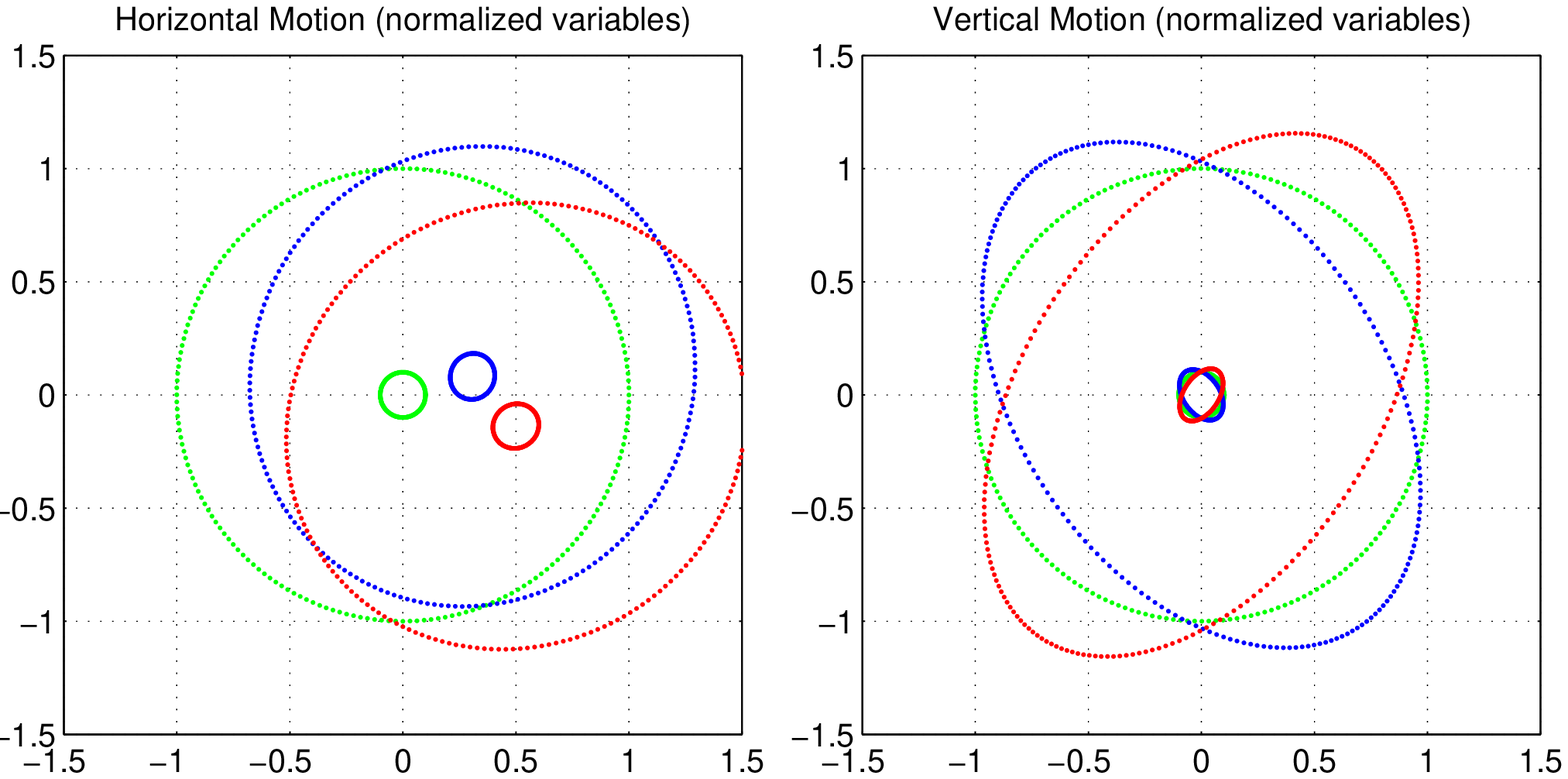}
    \vspace{-0.45cm}
    \caption{Phase space portraits of monochromatic 
    $0.1\sigma_{x,y}$ and $1\sigma_{x,y}$ ellipses
    (matched at the entrance) after tracking through
    the dogleg variant three. 
    The relative energy deviations are equal to
    $\pm 3\%$ and $\pm 5\%$.}
    \vspace{-0.3cm}
    \label{fig8}
\end{figure}

\end{document}